\UseRawInputEncoding
\documentclass[superscriptaddress, notitlepage, reprint,twocolumn]{revtex4-1}
\usepackage[english]{babel}
\usepackage{amsmath,amsthm,amssymb,lipsum}
\usepackage{amsfonts}
\usepackage[pdfborder={0 0 0}, colorlinks=true, urlcolor=blue, linkcolor=blue, citecolor=blue]{hyperref}
\usepackage{color}
\usepackage{graphicx}
\usepackage{float}
\usepackage{subfigure}
\usepackage{lipsum}
\usepackage{epstopdf}
\usepackage{dcolumn}
\usepackage{mathrsfs}
\usepackage[english]{babel}
\usepackage{csquotes}

\begin{document}

\title{Super-Heisenberg Scaling Using Nonlinear Quantum Scrambling }
\author{Dong  Xie}
\email{xiedong@mail.ustc.edu.cn}
\author{Chunling Xu}

\affiliation{College of Science, Guilin University of Aerospace Technology, Guilin, Guangxi 541004, People's Republic of China}

\begin{abstract}
Super-Heisenberg scaling, which scales as $N^{-\beta}$ with $\beta>1$ in terms of the number of particles $N$ or $T^{-\beta}$ in terms of the evolution time $T$, is better than Heisenberg scaling in quantum metrology. It has been proven that super-Heisenberg scaling can be achieved when the Hamiltonian of the system involves many-body interactions or the time-dependent terms. We demonstrate that nonlinear quantum scrambling facilitates the achievement of super-Heisenberg scaling $T^{-\beta}$ when the generator of the parameter is time-independent. More importantly, in dissipative systems, we can still obtain super-Heisenberg scaling in the  friction model. In the optical cavity system, an exponential improvement in measurement precision over time can be achieved by combining injected external squeezing and intracavity squeezing. Our work provides an optimal method for leveraging nonlinear resources to enhance the measurement precision of the driving field.
\end{abstract}
\maketitle

{\textit{Introduction}.-}By utilizing entangled quantum states of a linear system, the measurement precision limit of parameters can be improved from the standard quantum limit to the Heisenberg limit in terms of the number of particles $N$~\cite{lab1,lab2,lab3}.
It has been widely applied in many fields, such as, gravitational-wave detectors~\cite{lab4,lab5}, atomic spectroscopy~\cite{lab6,lab7,lab8,lab9,lab10,lab11}, precision tests of the fundamental laws of physics~\cite{lab12}, and atomic clocks~\cite{lab13,lab14,lab15}.

 By using nonlinear interactions, the measurement precision limit can be further improved, so that the super-Heisenberg scaling can be obtained. In the earlier work, Roy and Braunstein~\cite{lab16} derived a $1/2^N$ scaling based on the requirement to couple
 the $N$ constituent qubits to a rank-$N$ tensor field. Although Boixo et al.~\cite{lab17} argued that the exponential
 scaling is unphysical, it is still possible to achieve the super-Heisenberg limit by using the $z$-body interaction~\cite{lab17,lab18}.
Subsequently, a large amount of work was devoted to exploring the utilization of nonlinearity in open systems to achieve the super-Heisenberg limit~\cite{lab19,lab20,lab21,lab22,lab23,lab24}. Recently, Lordi et al.~\cite{lab25} developed a framework for analyzing noisy nonlinear metrology and established sufficient
 conditions for nonlinear metrology to be useful in the
 presence of quantum noise. Most of the work focuses on super-Heisenberg scaling associated with the number of particles $N$.
When the Hamiltonian of the probe system is time-dependent, super-Heisenberg scaling $T^{-\beta}$ in terms of the evolution time $T$ can be easily obtained. However, a critical question remains: when the Hamiltonian of the probe system is independent of time, can the super-Heisenberg  scaling be achieved through nonlinearity?

Recently, Montenegro et al.~\cite{lab26} showed that nonlinearity can enhance frequency estimation by virtue of quantum scrambling, where  local quantum information is spread out to a larger Hilbert space. In this Letter, we demonstrate that by choosing a proper Hamiltonian of the probe system, super-Heisenberg  scaling can be achieved through nonlinear quantum scrambling. We will establish the measurement precision limits for the driving signal fields (e.g., electromagnetic fields), and the conditions for achieving the super-Heisenberg scaling. In the dissipative case, we consider two models: the friction model and the cavity model. In the friction model, super-Heisenberg  scaling can be  obtained directly. In the cavity model, however, super-Heisenberg  scaling will disappear due to the dissipation process. Importantly, we find that a combination of injected external squeezing and intracavity squeezing can help to recover the super-Heisenberg  scaling, and even achieve an exponential improvement in measurement precision.

{\textit{Model}.-}We consider that the Hamiltonian is described by ($\hbar=1$)
 \begin{align}
H=\omega a^\dagger a+\frac{\lambda}{\sqrt{2}} [ae^{i(\omega_dt+\varphi)}+a^\dag e^{-i(\omega_dt+\varphi)}]+H_n,
\end{align}
where $\omega$ is the frequency of the system, $a\ (a^\dagger)$ denotes annihilation (creation) operator obeying the commutation relation $[a,a^\dagger]=1$, $\lambda$ is the unknown
amplitude of the driving signal that we intend to estimate, $\omega_d$ is the driving frequency, $\varphi$ denotes a reference phase, and $H_n$ is the nonlinear term of the Hamiltonian.

In this work, we consider that the nonlinear Hamiltonian belongs to a polynomial case, which is described by
 \begin{align}
H_n=G[\frac{ae^{i(\omega_{d}t+\vartheta)}+a^\dag e^{-i(\omega_{d}t+\vartheta)}}{\sqrt{2}}]^M,
\end{align}
where $G$ denotes nonlinear strength, $\vartheta$ denotes a reference phase, and the exponent $M$ is an integer greater than or equal to 2, i.e., $M\geq2$.
The nonlinear Hamiltonian  can be realized in microwave superconducting systems~\cite{lab27,lab28}.

Let $\omega_d=\omega$, $\varphi=0$ and $\vartheta=-\pi/2$, in the rotating reference frame, the Hamiltonian is rewritten as

 \begin{align}
H=\lambda X+G P^M,
\end{align}
where the momentum and position quadrature operators are defined as:  $P=\frac{1}{\sqrt{2i}}(a-a^\dagger)$, and $X=\frac{1}{\sqrt{2}}(a+a^\dagger)$.

{\textit{Quantum Fisher Information}.-}The quantum Cram\'{e}r-Rao theorem~\cite{lab29,lab30,lab31} has provided the lower bound of the measurement uncertainty $\delta \lambda$ for the unknown parameter $\lambda$,
 \begin{align}
\delta \lambda\geq[\nu \mathcal{F}(\lambda)]^{-1/2},
\end{align}
where $\nu$ represents the number of repeated measurements in the experiment, $\mathcal{F}(\lambda)$ is the quantum Fisher information with respect to $\lambda$.
For the pure state $|\psi(T)\rangle=e^{-iHT}|\psi_0\rangle$ at time $T$, the quantum Fisher information can be expressed as
 \begin{align}
\mathcal{F}(\lambda)=4(\langle\partial_\lambda\psi|\partial_\lambda\psi\rangle-|\langle\psi|\partial_\lambda\psi\rangle|^2),
\end{align}
where the abbreviation $\partial_\lambda=\frac{\partial}{\partial\lambda}$.
When $G$ is independent of $\lambda$ and for a long time  $T\gg1$, we can obtain that
 \begin{align}
\mathcal{F}(\lambda)\approx 4(G \lambda^{M-2} T^M)^2\Delta^2 P,\label{eq:6}
\end{align}
where the variance $\Delta^2 P=\langle\psi_0| P^2|\psi_0\rangle-\langle\psi_0| P|\psi_0\rangle^2.$
This result shows that the super-Heisenberg scaling $T^{-M}$ can be obtained by the nonlinear Hamiltonian.

When $G$ is proportional to $\lambda$ (i.e., $G=\xi \lambda$), more information about $\lambda$ has been encoded from the Hamiltonian. Intuitively, one would expect to obtain more information to estimate the parameter $\lambda$.
However, after direct calculation, we obtain the quantum Fisher information
 \begin{align}
\mathcal{F}&(\lambda)=\nonumber\\
&4[\langle\psi_0|(XT+gP^MT)^2|\psi_0\rangle-|\langle\psi_0|XT+gP^MT|\psi_0\rangle|^2]\nonumber\\
&\propto4T^2.
\end{align}
Obviously, this result is counter-intuitive, and fails to achieve the so-called \textquotedblleft super-Heisenberg scaling\textquotedblright.

To conduct a more in-depth analysis of the impact of brought about by the association between $G$ and $\lambda$, we consider that both $G$ and $\lambda$ are functions of a common parameter $y$. The quantum Fisher information with respect to $y$ is given by
 \begin{align}
\mathcal{F}(\lambda)\approx4[(G\lambda'-G'\lambda)\lambda^{M-2}T^M]^2\langle\Delta^2P\rangle,
\end{align}
where the partial derivatives are defined as $G'=\partial_yG$ and $\lambda'=\partial_y\lambda$.
The condition for achieving the super-Heisenberg scaling is
 \begin{align}
G\lambda'-G'\lambda\neq0.
\end{align}
The direct result of the derivation is $G\neq\xi \lambda$.
Therefore, as long as $G$ is not exactly proportional to $\lambda$, the super-Heisenberg scaling can be obtained.

{\textit{Optimal measurement operator}.-}With a specific measurement operator $M_e$, the measurement uncertainty of $\lambda$ is calculated by the error propagation formula:
 \begin{align}
\delta \lambda =\frac{\sqrt{\langle M_e^2\rangle-\langle M_e\rangle^2}}{|\partial_\lambda \langle M_e\rangle|},
\end{align}
where, without loss of generality, we have set the number of experimental repetitions to 1, i.e., $\nu=1$.
We consider that the measurement operator for $\lambda$ near the value $\lambda_c$ is
 \begin{align}
M_e=X+G/\lambda_cP^M-G/\lambda_c(P+\lambda_cT)^M.\label{eq:11}
\end{align}
According to the error-propagation formula, the measurement uncertainty of $\lambda\rightarrow\lambda_c$ is given by
\begin{align}
\delta\lambda&=\frac{\langle\delta X\rangle}{G\lambda^{M-2}t^M}\\
&\geq \frac{1}{2G\lambda_c^{M-2}t^M\langle\delta P\rangle}\\
&=\frac{1}{\sqrt{\mathcal{F}(\lambda_c)}},
\end{align}
where we have applied the Heisenberg uncertainty principle $\langle\delta X\rangle\geq1/\langle\delta P\rangle$.
This result is consistent with the previous result derived from the quantum Fisher information; therefore, the optimal measurement operator is $M_e$ as shown in Eq.~(\ref{eq:11}).

\begin{figure}[h]
\includegraphics[scale=0.32]{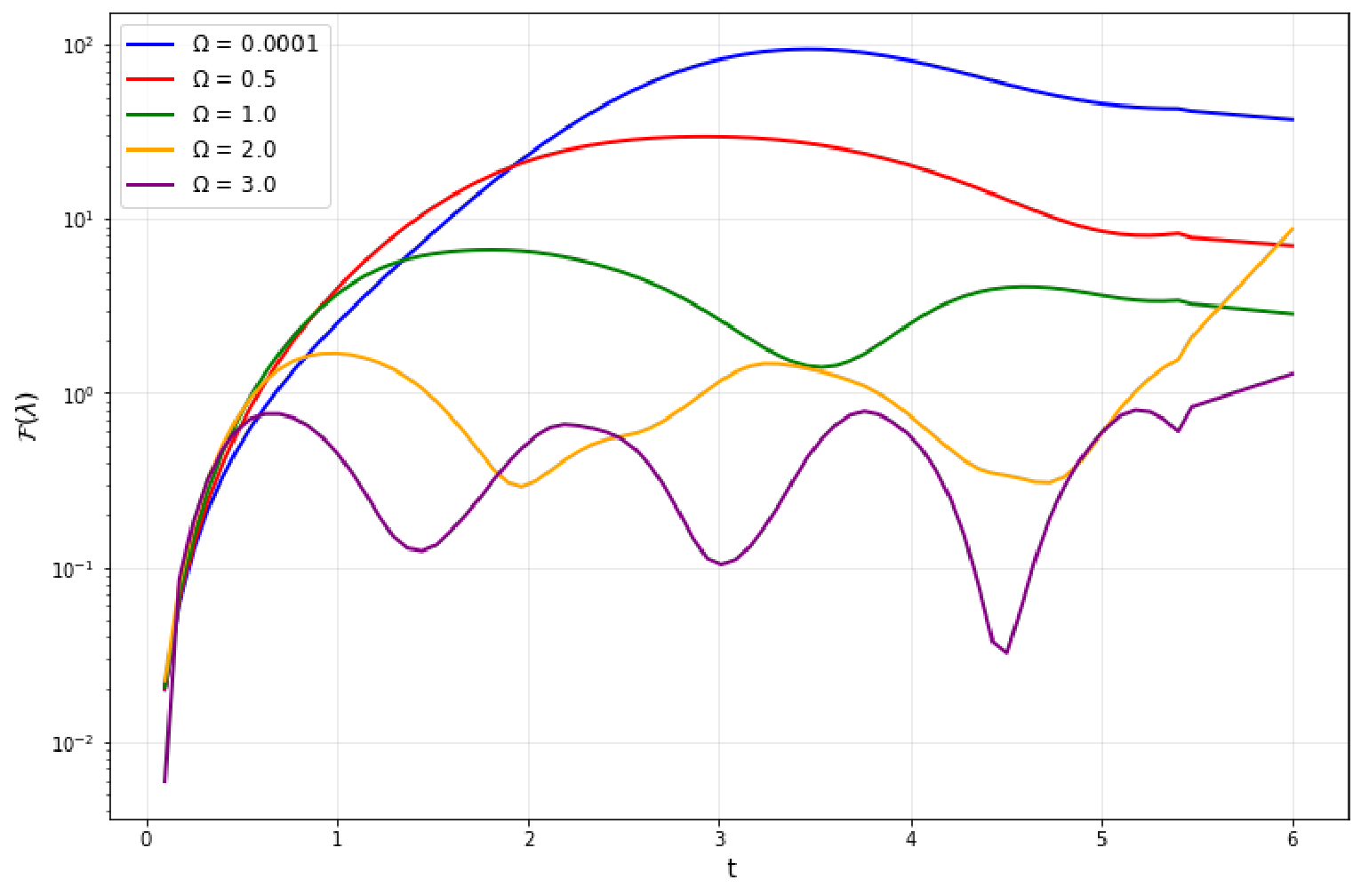}
 \caption{\label{fig.1}The time-evolution diagram of the quantum Fisher information of the parameter $\lambda$ with respect to different detunings $\Omega$. The dimensionless parameters have been set as: $M = 4, \alpha = 0.1, G = 0.1.$}
\end{figure}

{\textit{Influence of detuning}.-} When there is detuning, i.e., $\Omega=\omega-\omega_d$,
the evolution equations of the position operator and the momentum operator are:
 \begin{align}
 \dot{X}=M G P^{M-1}+\Omega P,\\
  \dot{P}=-\lambda- \Omega X.
\end{align}
When $M=2$, the analytical solutions of the above euqations at time $T$ are given by
 \begin{align}
 X&=\cos[ \sqrt{\Omega(2G+\omega)}T]X_0+\lambda\frac{\cos[ \sqrt{\omega(2G+\Omega)}T]-1}{\Omega}\nonumber\\
 &+\frac{\sin [\sqrt{\Omega(2G+\Omega)}T]\sqrt{2G+\Omega}}{\sqrt{\Omega}}P_0,\\
 P&=\frac{-\sqrt{\Omega}\sin [\sqrt{\Omega(2G+\Omega)}T]}{\sqrt{2G+\Omega}}X_0+\cos [\sqrt{\Omega(2G+\Omega)}T]P_0\nonumber\\
 &-\lambda\frac{\sin [\sqrt{\Omega(2G+\Omega)}T]}{\sqrt{(2G+\Omega)\Omega}}.
\end{align}
When $\sqrt{2\Omega(2G+\Omega) }T\ll1$, the corresponding uncertainty relation is given by
 \begin{align}
 \delta\lambda=\frac{\langle\delta X\rangle}{GT^2}.
\end{align}
However, as time elapses, the super-Heisenberg scaling will gradually be lost. This is because $X$ and $P$ are periodic, which means that the measurement information about $\lambda$ does not significantly increase over time. As shown in Fig.~\ref{fig.1}, it can be observed that when the time exceeds a certain value, the quantum Fisher information does not increase significantly. Additionally, as the detuning $\Omega$ increases, the quantum Fisher information quickly returns to a periodic state. Therefore, it is extremely necessary to reduce the detuning.
In order to resist the effects of the detuning, a squeezing process (e.g., spontaneous parametric down-conversion~\cite{lab32,lab33}) can be used, and the Hamiltonian of this squeezing process is expressed as
 \begin{align}
H_s=\chi ( a^2+a^{\dagger2}),
\end{align}
where $\chi$ denotes the coupling constant, which can be tuned by the pump.
Let $\chi=-\Omega/2$, the total Hamiltonian can be rewritten as
 \begin{align}
H_t=H+H_s=\lambda X+\Omega P^2+G P^M.
\end{align}
Dominated by the above Hamiltonian, the result in Eq.~(\ref{eq:6}) is recovered, which indicates that the squeezing process can effectively eliminate the impact induced by the detuning.

{\textit{Dissipation in the friction model}.-}
When the system suffers from dissipation due to interaction with noise, the dynamics can be described by Langevin equations.
In the friction model, the evolution of the position and momentum operators are given by~\cite{lab34}
 \begin{align}
 \partial_tX=i[H,X]+\eta_x(t),\\
  \partial_tP=i[H,P]-\gamma P+\eta_p(t),
\end{align}
where the noise $\eta_x(t)$ and $\eta_p(t)$ are operator-valued centred random variables, and $\gamma$ is the dissipation rate. The non-vanishing noise correlators are
 \begin{align}
\langle[\eta_x(t), \eta_p(t')]\rangle=i\gamma \delta(t-t'),\\
\langle\eta_y(t)\eta_p(t')\rangle=\gamma \coth(\frac{\omega}{T_e}) \delta(t-t'),
\end{align}
where $T_e$ denotes the temperature of the noise.
When $T\gg1$, the expectation values of the position and momentum operators are
 \begin{align}
\langle X\rangle\propto T,\\
\langle P\rangle=-\frac{\lambda}{\gamma}.
\end{align}
It is impossible to the super-Heisenberg scaling from the above measurement operators because the expectation values of position and momentum operators are not proportional to $T^M$.

We consider the Hamiltonian resulting from the exchange of the position operator $X$ and the momentum operator $P$, which is described by
 \begin{align}
H_p=\lambda P+G X^M.
\end{align}
After re-calculation, we obtained the expectation values of $P$ and $\delta^2P$ for $T\gg1$,
 \begin{align}
  {\langle P\rangle}&\approx\frac{M G \lambda^{M-1}T^{M-1}}{\gamma},\\
\langle\delta^2P\rangle&\approx\frac{1}{2}.
\end{align}
According to the error propagation formula, the measurement uncertainty $\delta \lambda_f$ of the parameter $\lambda$ in the frictional model is given by
\begin{align}
\delta \lambda_f=\frac{\gamma}{\sqrt{2}M(M-1) G \lambda^{M-2}T^{M-1}}.
\end{align}

This result recovers the super-Heisenberg scaling when $M>2$. Compared with the dissipation-free case in Eq.~(\ref{eq:6}), the measurement precision  merely loses a multiple of the measurement time, i.e., $\frac{\delta \lambda_f}{\delta \lambda}\propto T$.
Unlike the previous requirement for nonlinear measurement, only the momentum needs to be directly measured here, thus reducing the complexity of the measurement.

{\textit{Dissipation in the cavity model}.-}In the cavity model, the dynamics of the cavity model $a$ is described by the Langevin equation
 \begin{align}
 \partial_ta=i[H,a]-\gamma a+\sqrt{2\gamma}a_{in}(t),
\end{align}
where the cavity suffers from a squeezed vacuum reservoir with the noise operator $a_{in}(t)$. The correlations for the noise operator satisfy the following relations~\cite{lab35,lab36}
 \begin{align}
 \langle a^\dagger_{in}(t)a_{in}(t')\rangle=\sinh^2(r)\delta(t-t'),\\
  \langle a_{in}(t)a^\dagger_{in}(t')\rangle=\cosh^2(r)\delta(t-t'),\\
   \langle a_{in}(t)a_{in}(t')\rangle=-\frac{1}{2}\sinh(2r)\delta(t-t'),\\
    \langle a^\dagger_{in}(t)a^\dagger_{in}(t')\rangle=-\frac{1}{2}\sinh(2r)\delta(t-t'),
\end{align}
where $r$ denotes the squeezing parameter of the input noise.
Switching to the position and momentum operators, we can obtain the corresponding Langevin equations,
 \begin{align}
 \partial_tX=i[H,X]-\gamma X+A_x(t),\\
  \partial_tP=i[H,P]-\gamma P+A_p(t),
\end{align}
where the noise operators are $A_x(t)=\sqrt{\gamma}[{a_{in}(t)+a^\dagger_{in}(t)}]$, and $A_p(t)=-\sqrt{\gamma}i[{a_{in}(t)-a^\dagger_{in}(t)}]$.
Since both the momentum and position operators decay over time, it is impossible to directly obtain the super-Heisenberg scaling as in the friction model.
In order to obtain the super-Heisenberg scaling, a two-photon driving (i.e., squeezed case) can be employed to resist the dissipation rate.  The Hamiltonian for the two-photon driving is given by~\cite{lab37}
 \begin{align}
H_d=i\mu(a^{\dagger2}-a^2),
\end{align}
where $\mu$ denotes the driving strength.

When $\mu'=\mu-\gamma=0$, the expectation value of the measurement operator $P$ is given by
 \begin{align}
\langle P\rangle&\approx\frac{ -MG\lambda^{M-1}T^{M-1}}{2\gamma}.
  \end{align}
  When the squeezing parameter $r$ satisfies
  \begin{align}
e^{2r}=\frac{M(M-1)G\lambda^{M-2}T^{M-3/2}}{\sqrt{\gamma}},
  \end{align}
  the variance of the momentum operator reaches its minimum,
   \begin{align}
\langle \delta^2P\rangle\geq \frac{M(M-1)G\lambda^{M-2}T^{M-3/2}}{2\sqrt{\gamma}}.
  \end{align}
At this point, the minimum measurement uncertainty of the parameter $\lambda$  is
     \begin{align}
\delta \lambda=\frac{\sqrt{2}\gamma^{3/4}}{\sqrt{M(M-1)G\lambda^{M-2}}T^{(M-2)/4}}.
  \end{align}
From the above equation, it can be seen that the super-Heisenberg scaling can be obtained when $M>2$. This indicates that the combination of injected external squeezing and intracavity squeezing can counteract the dissipation process in the cavity model.

When $\mu>\gamma$ and $T\gg1$, we can obtain the expectation value of the measurement operator $P$ as follows:
 \begin{align}
 \langle P\rangle&\approx\frac{-MG(\lambda/\mu_-+\langle X_0\rangle)^{M-1}\exp[(M-1)(\mu_- T)]}{\mu_+}.
\end{align}
 When the squeezing parameter is given by
 \begin{align}
e^{2r}=\frac{[M(M-1)G(\langle X_0\rangle+\lambda/\mu_-)^{M-2}] e^{(M-1)\mu_-T}}{\sqrt{\mu_-{\mu_+}}},
  \end{align}
the optimal measurement uncertainty of  $\lambda$ is
 \begin{align}
\delta\lambda\geq\frac{\gamma^{1/2}\mu_+^{1/4}\mu_-^{3/4}}{\sqrt{M(M-1)G}(\lambda/\mu_-+\langle X_0\rangle)^{(M-2)/2}e^{(M-1)\mu_-T/2}}.
  \end{align}
Based on the above equation, we can see that when $M=2$, the measurement precision is independent of the initial position state. When $M> 2$, increasing the initial expected value of $X$ can improve the measurement precision. This is mainly achieved through higher-order nonlinearity, which amplifies the information of parameter $\lambda$ in the expectation value of $P$.
What's more, the above equation demonstrates that when $M$ is greater than or equal to 2, the measurement uncertainty of the parameters decreases exponentially over time. At this point, the amplification factor of the signal is much greater than that of the noise. This is achieved through the application of nonlinear quantum scrambling with the assistance of injected external squeezing and intracavity squeezing.

{\textit{Conclusion}.-}We have conducted a thorough investigation into the role of nonlinear quantum scrambling in enhancing parameter precision, and determined the theoretical limit of the achievable measurement precision. In a closed system, the super-Heisenberg scaling can be achieved under general conditions (i.e., when the nonlinear coupling coefficient $G$ is independent of the linear coupling coefficient $\lambda$ to be measured). An interesting point is that when the nonlinear coupling coefficient is proportional to the linear coupling coefficient being measured, it actually causes the super-Heisenberg scaling to vanish. The optimal measurement operator is obtained, and it is  proven that the optimal measurement precision can be achieved. An appropriate squeezing process can be employed to resist the effects of detuning between the frequency of the free Hamiltonian and the driving frequency. In the dissipative friction model, super-Heisenberg scaling can still be obtained by using a dissipation-free position operator to encode information.
In the dissipative cavity model, the measurement precision can be enhanced exponentially over time through the combined use of injected external squeezing and intracavity squeezing.
In both dissipative systems, measuring only the momentum is sufficient to obtain the super-Heisenberg scaling. Our results have opened up a pathway for the effective utilization of nonlinear quantum scrambling to enhance the measurement precision.

{\textit{Acknowledgements}.-}This research was supported by the National Natural Science Foundation of China (Grant No. 12365001 and No. 62001134), the Bagui youth top talent training program, and Guangxi Natural Science Foundation (Grant No. 2020GXNSFAA159047).

\clearpage

\subsection*{Supplemental Material: \textquotedblleft Super-Heisenberg Scaling Using Nonlinear Quantum Scrambling\textquotedblright }
In this Supplementary Material, we first give details about the derivation. In Sec. S1-S3, we provide the specific derivation processes of the quantum Fisher information  under  three scenarios: when $G$ is independent of $\lambda$, when $G$ is related to $\lambda$, and when both $G$ and $\lambda$ are functions of the measured variable $y$. An optimal measurement operator in the closed system is derived, and it is proven to be exactly equal to the precision limit value obtained from the quantum Fisher information. in Sec. S4. In Sec. S5, the influence of detuning has been discussed. The measurement uncertainty is meticulously derived in the dissipative friction model and cavity model in Sec. S6-S7.

\subsection*{S1. Calculation of Quantum Fisher Information: $G$ and $\lambda$ are independent of each other}
We start the calculation based on the following Hamiltonian
 \begin{align}
H=\lambda X+G P^M.
\tag{S1}
\end{align}
The unitary evolution dominated by the above Hamiltonian is given by
 \begin{align}
\exp(-iHT)&=\exp(-iGP^MT)\exp(-i\lambda XT)\nonumber\\
&\times\exp[\sum_{n=2}^{M+1}iG(-\lambda)^{n-1} C_n],\tag{S2}
\end{align}
where the terms $C_n$ are~\cite{s1}
 \begin{align}
C_n=\frac{M!(n-1)}{n!(M-n+1)!}P^{M-n+1}T^n.\tag{S3}
\end{align}
The unitary transformation of the position operator $X$  with respect to the last item above is
 \begin{align}
&\exp[\sum_{n=2}^{M+1}-iG(-\lambda)^{n-1} C_n]X\exp[\sum_{n=2}^{M+1}iG(-\lambda)^{n-1} C_n]\nonumber\\
&=X+[\sum_{n=2}^{M+1}-iG(-\lambda)^{n-1} C_n,X]\tag{S4}\\
&=X-\sum_{n=2}^{M}G(-\lambda)^{n-1}\frac{M!(n-1)}{n!(M-n)!}P^{M-n}T^n.\tag{S5}
\end{align}

At time $T$, the pure state is described by
\begin{align}
|\psi(T)\rangle=\exp(-iHT)|\psi_0\rangle,\tag{S6}
\end{align}
where $|\psi_0\rangle$ denotes the initial state of the system. Utilizing the above equations, the quantum Fisher information with respect to the parameter $\lambda$ can be derived
 \begin{align}
&\mathcal{F}_\lambda=4(\langle\partial_\lambda\psi(T)|\partial_\lambda\psi(T)\rangle-|\langle\psi|\partial_\lambda\psi(T)\rangle|^2)\tag{S7}\\
&=4\langle\psi_0|\exp[\sum_{n=2}^{M+1}-iG(-\lambda)^{n-1} C_n][ X T+\sum_{n=2}^{M+1}G(n-1)\nonumber\\
&\times(-\lambda)^{n-2} C_n]^2
\exp[\sum_{n=2}^{M+1}+iG(-\lambda)^{n-1} C_n]|\psi_0\rangle\nonumber\\
&-4|\langle\psi_0|\exp[\sum_{n=2}^{M+1}-iG(-\lambda)^{n-1} C_n][ X T+\sum_{n=2}^{M+1}g(n-1)\nonumber\\
&\times(-\lambda)^{n-2} C_n]\exp[\sum_{n=2}^{M+1}+iG(-\lambda)^{n-1} C_n]|\psi_0\rangle|^2\tag{S8}\\
&=4\langle\psi_0|[XT-\sum_{n=2}^{M}G(-\lambda)^{n-1}\frac{M!(n-1)}{n!(M-n)!}P^{M-n}T^{n+1}\nonumber\\
&+\sum_{n=2}^{M+1}G(n-1)(-\lambda)^{n-2}\frac{M!(n-1)}{n!(M-n+1)!}P^{M-n+1}T^n]^2|\psi_0\rangle\nonumber\\
&-4|\langle\psi_0|[XT-\sum_{n=2}^{M}G(-\lambda)^{n-1}\frac{M!(n-1)}{n!(M-n)!}P^{M-n}T^{n+1}\nonumber\\
&+\sum_{n=2}^{M+1}G(n-1)(-\lambda)^{n-2}\frac{M!(n-1)}{n!(M-n+1)!}P^{M-n+1}T^n]|\psi_0\rangle|^2\tag{S9}\\
&\approx 4\langle\psi_0|(XT+PG\lambda^{M-2} T^M)^2|\psi_0\rangle\nonumber\\
&\ \ -4|\langle\psi_0|XT+PG\lambda^{M-2} T^M|\psi_0\rangle|^2\tag{S10}\\
&\approx 4(G \lambda^{M-2} T^M)^2\Delta^2 P.\tag{S11}
\end{align}

\subsection*{S2. Calculation of Quantum Fisher Information: $G$ and $\lambda$ are relevant}
When $G\propto \lambda$, i.e., $G=\xi\lambda$, the quantum Fisher information is directly given by
 \begin{align}
\mathcal{F}_\lambda&=4[\langle\psi_0|(XT+GP^MT)^2|\psi_0\rangle-|\langle\psi_0|XT+GP^MT|\psi_0\rangle|^2]\tag{S12}\\
&\propto4T^2.\tag{S13}
\end{align}
The unitary evolution can also rewritten as
\begin{align}
&\exp(-iHT)=\exp(-i\lambda XT)\exp(-iV),\tag{S14}\\
&V=GP^MT+\sum_{n=2}^{M+1} G\frac{M!(-\lambda)^{n-1}}{n!(M-n+1)!}P^{M-n+1}T^n.\tag{S15}
\end{align}
The unitary transformation of the position operator $X$  with respect to the operator $V$ is
 \begin{align}
&\exp(iV)X\exp(-iV)=X+i[V,X]\tag{S16}\\
&=X+MGP^{M-1}T+\sum_{n=2}^{M} G\frac{M!(-\lambda)^{n-1}}{n!(M-n)!}P^{M-n}T^n.\tag{S17}
\end{align}

When $G$  is a function of $\lambda$, the quantum Fisher information of $\lambda$ can be calculated by the following process
 \begin{align}
&\mathcal{F}_\lambda=4(\langle\partial_\lambda\psi(T)|\partial_\lambda\psi(T)\rangle-|\langle\psi|\partial_\lambda\psi(T)\rangle|^2)\tag{S18}
\end{align}
 \begin{align}
&=4\langle \{MGP^{M-1}T^2+\sum_{n=2}^{M} \frac{GM!(-\lambda)^{n-1}}{n!(M-n)!}P^{M-n}T^{n+1}\nonumber\\
&+XT+\sum_{n=2}^{M+1} [G'(-\lambda)^{n-1}-(n-1)G(-\lambda)^{n-2}]\nonumber\\
&\times\frac{M!}{n!(M-n+1)!}P^{M-n+1}T^{n}+MG'P^{M-1}T\}^2\rangle\nonumber\\
&-4|\langle MGP^{M-1}T^2+\sum_{n=2}^{M} \frac{GM!(-\lambda)^{n-1}}{n!(M-n)!}P^{M-n}T^{n+1}\nonumber\\
&+XT+\sum_{n=2}^{M+1} [G'(-\lambda)^{n-1}-(n-1)G(-\lambda)^{n-2}]\nonumber\\
&\times\frac{M!}{n!(M-n+1)!}P^{M-n+1}T^{n}+MG'P^{M-1}T\rangle|^2\tag{S19}
\end{align}
 \begin{align}
&=4\langle [XT+GP^{M-1}T^2(G-G'\lambda)/2+MG'P^{M-1}T\nonumber\\
&+\sum_{n=3}^{M+1}(G-G'\lambda)(-\lambda)^{n-2}\frac{M!}{n!(M-n+1)!}P^{M-n+1}T^{n}]^2\rangle\nonumber\\
&-4|\langle [XT+MGP^{M-1}T^2(G-G'\lambda)/2+MG'P^{M-1}T\nonumber\\
&+\sum_{n=3}^{M+1}(G-G'\lambda)(-\lambda)^{n-2}\frac{M!}{n!(M-n+1)!}P^{M-n+1}T^{n}]\rangle|^2,\tag{S20}
\end{align}
where $G'=\partial_\lambda G$.
From the above equation, one can clearly see that the coefficient of the higher-order terms of time ($G-G'\lambda$) is exactly cancelled out when $G$ is proportional to $\lambda$: $G=\xi \lambda$. When $ G'>0$, the coefficient $G-G'\lambda<0$. At this point, if the slope of $G$ with respect to $\lambda$ is greater than 0, the quantum Fisher information of the system can be reduced. Conversely, the quantum Fisher information about $\lambda$ can
be improved to a certain extent, but it does not affect the exponential relationship with time.

\subsection*{S3. Calculation of Quantum Fisher Information: $G$ and $\lambda$ are functions of $y$}
When both $G$ and $\lambda$ are functions of $y$,  the quantum Fisher information with respect to $y$ can be derived by the following process
 \begin{align}
&\mathcal{F}_\lambda=4(\langle\partial_\lambda\psi(T)|\partial_\lambda\psi(T)\rangle-|\langle\psi|\partial_\lambda\psi(T)\rangle|^2)\tag{S21}
\end{align}
 \begin{align}
&=4\langle \{\lambda'XT+\sum_{n=2}^{M} \lambda'G(-\lambda)^{n-1}\frac{M!}{n!(M-n)!}P^{M-n}T^{n+1}\nonumber\\
&+MG'P^{M-1}T+\sum_{n=2}^{M+1} [G'(-\lambda)^{n-1}-(n-1)\lambda'G(-\lambda)^{n-2}]\nonumber\\
&\times\frac{M!}{n!(M-n+1)!}P^{M-n+1}T^{n}+\lambda'MGP^{M-1}T^2\}^2\rangle\nonumber\\
&-4|\langle \lambda'XT+\sum_{n=2}^{M} \lambda'G(-\lambda)^{n-1}\frac{M!}{n!(M-n)!}P^{M-n}T^{n+1}\nonumber\\
&+MG'P^{M-1}T+\sum_{n=2}^{M+1} [G'(-\lambda)^{n-1}-(n-1)\lambda'G(-\lambda)^{n-2}]\nonumber\\
&\times\frac{M!}{n!(M-n+1)!}P^{M-n+1}T^{n}+\lambda'MGP^{M-1}T^2\rangle|^2\tag{S22}
\end{align}
 \begin{align}
&=4\langle [\lambda'XT+MP^{M-1}T^2(G\lambda'-G'\lambda)/2+MG'P^{M-1}T\nonumber\\
&+\sum_{n=3}^{M+1}(G\lambda'-G'\lambda)(-\lambda)^{n-2}\frac{M!}{n!(M-n+1)!}P^{M-n+1}T^{n}]^2\rangle\nonumber\\
&-|\langle [XT+MP^{M-1}T^2(G\lambda'-G'\lambda)/2+MG'P^{M-1}T\nonumber\\
&+\sum_{n=3}^{M+1}(G\lambda'-G'\lambda)(-\lambda)^{n-2}\frac{M!}{n!(M-n+1)!}P^{M-n+1}T^{n}]\rangle|^2,\tag{S23}
\end{align}
where $G'=\partial_yG$ and $\lambda'=\partial_y\lambda$.
When the condition $G\lambda'-G'\lambda\neq 0$ and for long time $T\gg1$, the quantum Fisher information can be reduced to the following form
 \begin{align}
\mathcal{F}_\lambda\approx 4[(G\lambda'-G'\lambda)(-\lambda)^{M-2}T^M]^2\langle\Delta^2P\rangle.\tag{S24}
\end{align}

\subsection*{S4. The optimal measurement operator}
The expectation value of the position operator $X$ is
 \begin{align}
\langle X\rangle&=\langle\psi_0|\exp[i(\lambda X+g P^M)T]X\exp[-i(\lambda X+g P^M)T]|\psi_0\rangle\tag{S25}\\
&=\langle X+MGP^{M-1}T+\sum_{n=2}^{M} \frac{M!(-\lambda)^{n-1}}{n!(M-n)!}G P^{M-n}T^n\rangle\tag{S26}\\
&=\langle X-G/\lambda(P-\lambda T)^M+G/\lambda P^M\rangle.\tag{S27}
\end{align}
The expectation value of the measurement operator $M_e$ is
 \begin{align}
\langle M_e\rangle&=\langle \psi|X+G/\lambda_cP^M-G/\lambda_c(P+\lambda_ct)^M|\psi\rangle \tag{S28}\\\
&=\langle\psi_0| X-G/\lambda(P-\lambda t)^M+G/\lambda_c(P-\lambda t)^M\nonumber\\\
&+G/\lambda P^M-G/\lambda_c [P+(\lambda_c-\lambda)t]^M|\psi_0\rangle.\tag{S29}\
\end{align}
According to the error propagation formula, the measurement uncertainty of  $\lambda$ is
 \begin{align}
\delta\lambda&=\frac{\sqrt{\langle M_e^2\rangle-\langle M_e\rangle^2}}{|\partial_\lambda\langle M_e\rangle|}\tag{S30}\\
&=\frac{\langle\delta X\rangle}{G\lambda^{M-2}t^M}\tag{S31}\\
&\geq \frac{1}{2G\lambda^{M-2}t^M\langle\delta P\rangle}=\frac{1}{\sqrt{\mathcal{F}(\lambda)}}.\tag{S32}
\end{align}

\subsection*{S5. The influence of detuning}
The detuning Hamiltonian is described by the following form
 \begin{align}
\Delta H=\Omega a^\dagger a,\tag{S33}
\end{align}
    where  the detuning $\Omega=\omega-\omega_d $.
The evolution equations of the position operator and the momentum operator are:
 \begin{align}
 \dot{X}&=M G P^{M-1}+\Omega P,\tag{S34}\\
  \dot{P}&=-\lambda- \Omega X.\tag{S35}
\end{align}
When $M=2$, the analytical solutions of the above equations are given by
 \begin{align}
 X&=\cos[ \sqrt{\Omega(2G+\omega)}T]X_0+\lambda\frac{\cos[ \sqrt{\Omega(2G+\Omega)}T]-1}{\Omega}\nonumber\\
 &+\frac{\sin [\sqrt{\Omega(2G+\omega)}T]\sqrt{2G+\Omega}}{\sqrt{\Omega}}P_0,\tag{S36}\\
 P&=\frac{-\sqrt{\Omega}\sin [\sqrt{\Omega(2G+\Omega)}T]}{\sqrt{2G+\Omega}}X_0+\cos [\sqrt{\Omega(2G+\Omega)}T]P_0\nonumber\\
 &-\lambda\frac{\sin [\sqrt{\Omega(2G+\Omega)}T]}{\sqrt{(2G+\Omega)\Omega}}.\tag{S37}
\end{align}
When $\sqrt{2\Omega(2G+\Omega) }T\ll1$, the optimal measurement operator is $M_e=X+\frac{G}{\lambda_c}P^2-\frac{G}{\lambda_c}(P+\lambda_cT)^2$.
The corresponding uncertain relationship is
 \begin{align}
 \delta\lambda=\frac{\langle\delta X\rangle}{GT^2}.\tag{S38}
\end{align}

\subsection*{S6. Dissipation in the friction model}
In the friction model, the quantum-Langevin eqautions are given by
 \begin{align}
 \dot{X}&=M G P^{M-1}+\eta_x,\tag{S39}\\
  \dot{P}&=-\lambda-\gamma P+\eta_p.\tag{S40}
\end{align}
As a result, the analytical forms of the momentum operator and the position operator are
 \begin{align}
  {P}&=\exp(-\gamma T)P_0+\int_{0}^Tdt'\exp[-\gamma (T-t')][-\lambda+\eta_p(t')],\tag{S41}\\
  X&=X_0+\int_{0}^Tdt'( M G P^{M-1}+\eta_x).\tag{S42}
\end{align}

When $\gamma T\gg1$, the above solutions can be further simplified as
 \begin{align}
  {P}&=-\lambda/\gamma+\int_{0}^Tdt'\exp[-\gamma (T-t')]\eta_p(t'),\tag{S43}\\
  X&=X_0+MG[-\lambda/\gamma+\int_{0}^\infty dt'\exp(-\gamma t')\eta_y(t')]^{M-1}T.\tag{S44}
\end{align}

By exchanging the positions of $X$ and $P$, the Hamiltonian is described by
 \begin{align}
H=\lambda P+G X^M.\tag{S45}
\end{align}
Under the control of this Hamiltonian, the corresponding quantum-Langevin equations are
 \begin{align}
 \dot{X}&=\lambda+\eta_x,\tag{S46}\\
  \dot{P}&=-MGX^{M-1}+\eta_p-\gamma P.\tag{S47}
\end{align}

By the same derivation, when $\gamma T\gg1$, the solutions at time $T$ are given by
 \begin{align}
 &{X}=X_0+\lambda T+\int_{0}^Tdt'\eta_x(t')\tag{S48}\\
  &{P}=\int_{0}^T dt'\exp(-\gamma t')\eta_p(t')-\nonumber\\
  &\int_{0}^Tdt'\exp[-\gamma(t- t')]MG[X_0+\lambda t'+\int_{0}^{t'} dt''\eta_x(t'')]^{M-1}.\tag{S49}
\end{align}

The expectation values are given by
 \begin{align}
  {\langle P\rangle}&=\int_{0}^Tdt'\exp[-\gamma(T- t')]MG(X_0+\lambda t')^{M-1}\tag{S50}\\
  &\approx\frac{M G \lambda^{M-1}T^{M-1}}{\gamma}.\tag{S51}
\end{align}
 \begin{align}
  &{\langle \delta ^2P\rangle}=\langle[\int_{0}^Tdt'\exp[-\gamma(T- t')]MG(X_0+\lambda t')^{M-1}]^2\rangle\nonumber\\
  &-|\langle\int_{0}^Tdt'\exp[-\gamma(T- t')]MG(X_0+\lambda t')^{M-1}\rangle|^2+\frac{1}{2}\tag{S52}\\
  &\approx\frac{1}{2}.\tag{S53}
\end{align}

\subsection*{S7. Dissipation in  the cavity model}
In the dissipative cavity model, the quantum-Langevin equations are
 \begin{align}
 \dot{X}&=-\gamma X+\lambda+A_x,\tag{S54}\\
  \dot{P}&=-MGX^{M-1}+A_p-\gamma P.\tag{S55}
\end{align}
The solutions is given by
 \begin{align}
 {X}&=\frac{\lambda}{\gamma}(1-\exp(-\gamma T))+\int_{0}^Tdt'\exp[-\gamma(T- t')]A_x(t')\nonumber\\
 &+\exp(-\gamma T)X_0,\tag{S56}\\
  {P}&=\exp(-\gamma T)P_0+\int_{0}^T dt'\exp(-\gamma t')A_p(t')\nonumber\\
  &-\int_{0}^Tdt'\exp[-\gamma(T- t')]MGX(t')^{M-1}.\tag{S57}
\end{align}
The expectation value of the momentum operator is given by
 \begin{align}
 \langle P\rangle=\frac{GM\lambda^{M-1}}{\gamma^M}.\tag{S58}
\end{align}

Including the two-photon driving, the quantum-Langevin equations are rewritten as
 \begin{align}
 \dot{X}&=(\mu-\gamma) X+\lambda+A_x,\tag{S59}\\
  \dot{P}&=-MGX^{M-1}+A_p+(-\mu-\gamma) P,\tag{S60}
\end{align}
where $\mu$ denotes the two-photon driving strength.
By solving the above equations, the general forms of the position and momentum operators at time $T$ are given by
 \begin{align}
 {X}=&\frac{\lambda}{\mu_-}[\exp(\mu_- T)-1]+\int_{0}^Tdt'\exp[\mu_-(T- t')]A_x(t')\nonumber\\
 &+\exp(\mu_- T)X_0,\tag{S61}\\
  {P}=&-\int_{0}^Tdt'\exp[-\mu_+(T- t')]MGX(t')^{M-1}\nonumber\\
  &+\int_{0}^Tdt'\exp(-\mu_+ t')A_p(t')+\exp(-\mu_+ T)P_0,\tag{S62}
\end{align}
where $\mu_\pm=\mu\pm\gamma$.

When $\mu=\gamma$ and $T\gg1$, the expectation values are given by
 \begin{align}
 \langle X\rangle&\approx \lambda T+ \langle X_0\rangle,\tag{S63}\\
\langle P\rangle&\approx\frac{ -MG\lambda^{M-1}T^{M-1}}{2\gamma}.\tag{S64}
  \end{align}
By using the correlations for the noise operator $a_{in}(t)$, we can obtain the correlations for $A_x(t)$ and $A_p(t)$
 \begin{align}
  \langle A_x(t)A_x(t')\rangle&=\gamma e^{-2r}\delta(t-t'),\tag{S65}\\
   \langle A_p(t)A_p(t')\rangle&=\gamma e^{2r}\delta(t-t').\tag{S66}
  \end{align}

We further obtain that the variance of the momentum operator $P$ is
 \begin{align}
  \langle \delta^2P\rangle&\approx \frac{(M-1)^2M^2G^2\lambda^{2(M-2)} T^{2(M-2)}}{(2\gamma)^2} \nonumber\\
  &\times\langle\int_0^Tdt' A_x(t')\int_0^Tdt'' A_x(t'')\rangle+\gamma e^{2r}/4\tag{S67}\\
&=\frac{(M-1)^2M^2G^2\lambda^{2(M-2)} T^{(2M-3)}e^{-2r}}{4\gamma}+ e^{2r}/4\tag{S68}\\
&\geq\frac{M(M-1)G\lambda^{M-2}T^{M-3/2}}{2\sqrt{\gamma}},\tag{S69}
  \end{align}
where the condition for setting the equality in the last equation is the squeezing parameter
 \begin{align}
e^{2r}=\frac{M(M-1)G\lambda^{M-2}T^{M-3/2}}{\sqrt{\gamma}}.\tag{S70}
  \end{align}
Based on the error propagation formula, the measurement uncertainty of  $\lambda$ is derived
 \begin{align}
\delta\lambda=\frac{ \sqrt{\langle \delta^2P\rangle}}{|\partial_\lambda\langle P\rangle |}\geq\frac{\sqrt{2}\gamma^{3/4}}{\sqrt{M(M-1)G\lambda^{M-2}}T^{(M-2)/4}}.\tag{S71}
  \end{align}

When $\mu>\gamma$ and $T\gg1$, we can obtain the expectation values
 \begin{align}
 \langle X\rangle&\approx(\lambda/\mu_-+\langle X_0\rangle)\exp(\mu_- T),\tag{S72}\\
 \langle P\rangle&\approx\frac{-MG(\lambda/\mu_-\exp(\mu_- T)+\exp(\mu_- T)\langle X_0\rangle)^{M-1}}{\mu_+}.\tag{S73}
\end{align}
Utilizing the the correlations of the noise, we further achieve the variance of the momentum operator
 \begin{align}
&\langle \delta^2P\rangle\approx\nonumber\\
&\frac{[M(M-1)G\langle X\rangle^{M-2}]^2\langle[\int_0^Tdt'e^{\mu_-(T-t')}A_x(t')]^2\rangle}{{\mu_+}^2}+\frac{\gamma e^{2r}}{2\mu_+}\tag{S74}\\
&=\frac{[M(M-1)G\langle X\rangle^{M-2}]^2\gamma e^{2\mu_-T-2r}}{2\mu_-{\mu_+}^2}+\frac{\gamma e^{2r}}{2\mu_+}\tag{S75}\\
&\geq \frac{[M(M-1)G(\langle X_0\rangle+\lambda/\mu_-)^{M-2}]{\gamma} e^{(M-1)\mu_-T}}{\mu_+\sqrt{\mu_-{\mu_+}}}.\tag{S76}
 \end{align}
where the condition for the equality to hold is that the squeezing parameter satisfies
 \begin{align}
e^{2r}=\frac{[M(M-1)G(\langle X_0\rangle+\lambda/\mu_-)^{M-2}] e^{(M-1)\mu_-T}}{\sqrt{\mu_-{\mu_+}}}.\tag{S77}
  \end{align}
Based on the error propagation formula, the measurement uncertainty of the parameter $\lambda$ is derived by the following equation
 \begin{align}
\delta\lambda&=\frac{ \sqrt{\langle \delta^2P\rangle}}{|\partial_\lambda\langle P\rangle |}\tag{S78}\\
&\geq\frac{\gamma^{1/2}\mu_+^{1/4}\mu_-^{3/4}}{\sqrt{M(M-1)G}(\lambda/\mu_-+\langle X_0\rangle)^{(M-2)/2}e^{(M-1)\mu_-T/2}}.\tag{S79}
  \end{align}

{}

\end{document}